\documentstyle[aps,multicol,epsfig]{revtex}
\begin{document}
% \draft command makes pacs numbers print
%\draft
\title{Magnetoresistance of RuO$_2$-based 
resistance thermometers below 0.3~K}  

% repeat the \author\address pair as needed
\author{Michio Watanabe,$^1$\footnote{Present address:  
Semiconductors Laboratory, RIKEN, 2-1 Hirosawa, Wako-shi, 
Saitama 351-0198, Japan} 
Masashi Morishita,$^2$ and Youiti Ootuka$^2$} 
\address{$^1$Dept. Applied Physics and Physico-Informatics, 
Keio University, Yokohama 223-8522, Japan\\
$^2$Institute of Physics, University of Tsukuba, 
1-1-1 Tennodai, Tsukuba, Ibaraki 305-8571, Japan  
}
\date{Received 11 July 2000; accepted 26 March 2001}
\maketitle
\begin{abstract}
We have determined the magnetoresistance of RuO$_2$-based resistors 
(Scientific Instruments (SI) RO-600) between 0.05 and 0.3~K 
in magnetic fields up to 8~T.  The magnetoresistance is negative 
around 0.5~T and then becomes positive at larger fields.  
The magnitude of the negative magnetoresistance increases rapidly 
as the temperature is lowered, while that of the positive 
magnetoresistance has smaller temperature dependence.  
We have also examined the temperature dependence of the resistance 
below 50~mK in zero magnetic field.  It is described in the context 
of variable-range-hopping (VRH) conduction down to 15~mK.  
Hence, the resistors can be used as thermometers down to at least 15~mK.  
\end{abstract}
% insert suggested PACS numbers in braces on next line
%\vspace*{5mm}
%\pacs{{\it Key words:} thermometry at very low temperatures; 
%thick-film chip resistors; 
%magnetoresistance; ruthenium oxide resistance thermometers
%}
\vspace*{6mm}
\begin{multicols}{2}
\section{Introduction}
As promising low-temperature thermometers, RuO$_2$-based thick-film 
chip resistors have been introduced (for example~\cite{Rub97}).  
The advantages of such resistors are reproducibility and 
reasonably small magnetoresistance.  
In order to implement the accuracy of the thermometry in magnetic fields, 
a lot of works has been devoted to the magnetoresistance measurements of
 RuO$_2$-based 
resistors~\cite{Kop83,Doi84,Bos86,Li86,Mei89,Wil90,Bri91,Uhl95,Nep96,Goo98}. 
Unfortunately, the magnetoresistance seems to be dependent on the detail of 
the manufacturing process.  In the case of commercial resisters, 
results vary even in the sign of the magnetoresistance depending 
upon the manufacturing company as summarized in [\ref{Bri91}, Table~1].  

Recently RuO$_2$-based resistors produced by Scientific Instruments 
Inc. (SI) are used in many laboratories.  However, there is no substantial 
publication on their magnetoresistance to the best of our knowledge.  
This is the motivation for our examining SI's resistors in this work.  
We focus on the temperature range of $T<0.3$~K, where thermometry 
based on a physical quantity that is nominally independent of the 
magnetic fields is extremely troublesome, and hence, information 
on the magnetoresistance of resistance thermometers is extremely valuable.  
Note that the vapor pressure of $^4$He or $^3$He is no longer appropriate 
in this temperature range for the purpose.  
We have determined the magnetoresistance at temperatures 
down to 0.05~K in fields up to 8~T.  

In addition to the magnetoresistance, 
we investigate the applicability of the resistors below 0.05~K.  
Concerning SI's resistors, the calibration is commercially available 
down to 0.05~K, while at present $^{3}$He-$^{4}$He dilution refrigerators 
with base temperatures of $0.02-0.03$~K are installed in many laboratories.  
Hence it should be a matter of great interest how one can describe 
the temperature dependence of the resistance, or how well 
the calibration table is extrapolated to lower temperatures.

\section{Experiment}
We measured the resistance of two RuO$_2$-based thick-film 
chip resistors (SI model RO-600A, S/N 1848 and 1849) 
at temperatures $T<0.3$~K 
using a $^{3}$He-$^{4}$He dilution refrigerator.  
Magnetic fields up to $B=8$~T were applied 
by means of a superconducting solenoid.  
The resistors are ``exposed" (not canned) chips.  
One of them [Resister~A (S/N~1848)] was placed so that the direction 
of the magnetic field was parallel to the film.  
For the other [Resistor~B (S/N~1849)], the direction 
of the magnetic field was perpendicular to the film, 
and hence, perpendicular to the current flow as well.  
The temperature was determined by a $^3$He-melting-curve 
thermometer (MCT)~\cite{Gre86}.  
The MCT was placed in a region where the magnetic field was always 
nominally zero, i.e., magnetic fields applied for the magnetoresistance 
measurements were compensated in the region.  
Moreover, the $^3$He melting curve is known to have a sufficiently small 
magnetic-field dependence in the temperature range 
of this work~\cite{Kra87}.  
Hence, the temperature determined by the MCT is not affected 
by the magnetic-field applied for the magnetoresistance 
measurements at all.
  
Both the resistors and the MCT were thermally connected 
to the mixing chamber of the refrigerator 
with thick cold fingers made of pure copper or pure silver.  
As for the resistors, the signal leads, which also act as 
a heat sink, were glued to one of the cold fingers 
with GE7031 varnish.   
For the resistance measurements we employed ac methods 
at $f\leq25$~Hz.  
The output voltage of the sample was detected 
by a voltage amplifier (DL Instruments 1201) and/or a lock-in 
amplifier (EG\&G Princeton Applied Research 124A 
or Standford Research System SR830DSP) at $T\leq0.09$~K.  
At $T\geq0.09$~K, we used a resistance bridge (RV-Elekroniikka AVS-45).  
We have confirmed that the resistance obtained by the two methods 
agree with each other at $T=0.09$~K.  

%%%%%%%%%%%%%%%%%%%%%%%%%%%%%%%%%%%%%%%%%%%%%
%%%%  Fig.1: deltaR             %%%%%%%%%%%%%
%%%%%%%%%%%%%%%%%%%%%%%%%%%%%%%%%%%%%%%%%%%%%
\begin{figure}
\centerline{
\psfig{
file=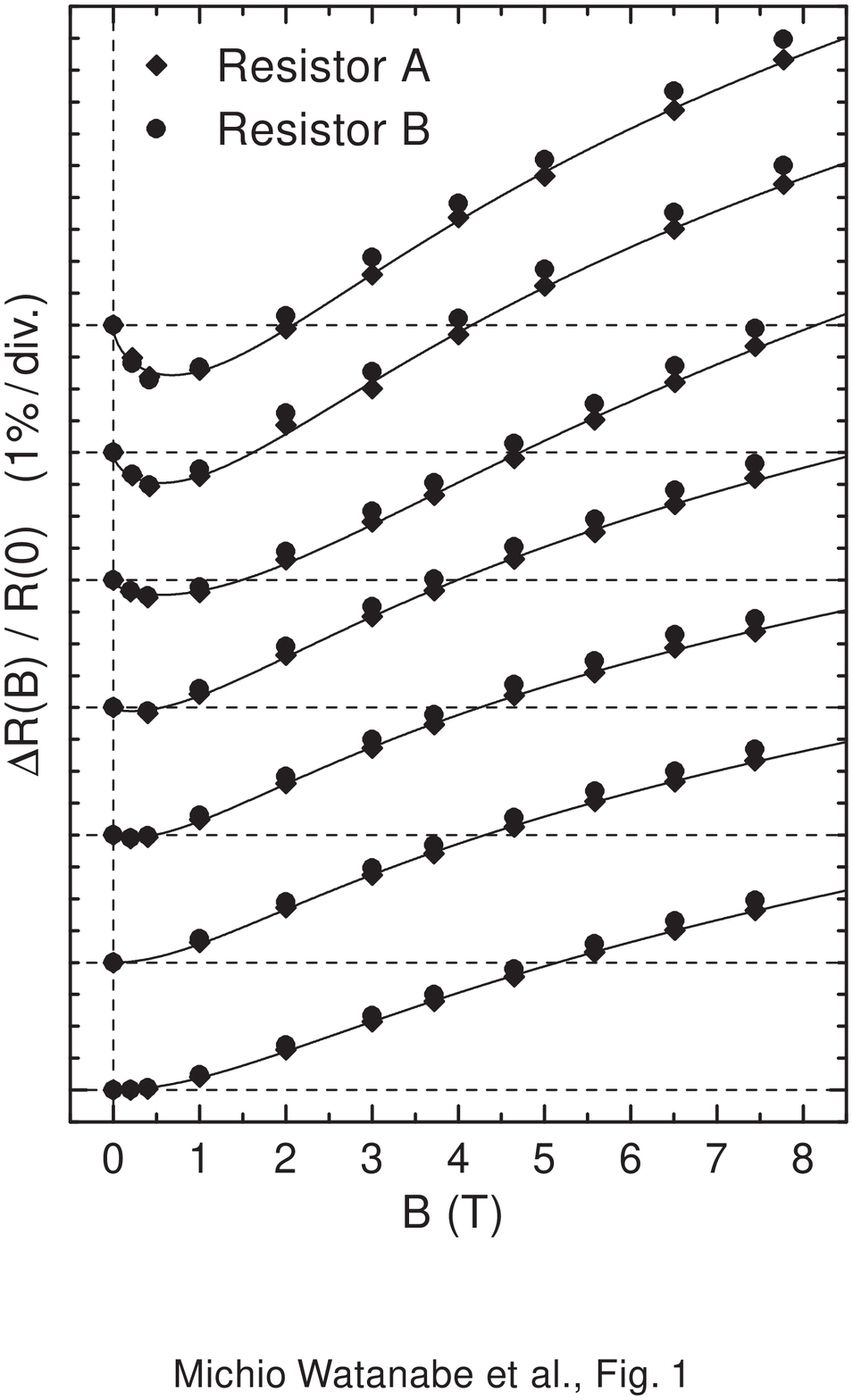,
width=.8\columnwidth,
bbllx=68pt,
bblly=131pt,
bburx=509pt,
bbury=783pt,clip=,
angle=0}
}
\caption{
Relative change of resistance as a function of magnetic field 
for the two RuO$_2$-based resistors at constant temperatures.  
The temperatures from top to bottom in units of K are 0.05, 0.06, 
0.09, 0.12, 0.16, 0.20, 0.24, respectively.  
The origin of each data set is offset for clarity.  
The solid curves represent the fits of $\Delta R(B)/R(0)=A_1B^{1/2} 
[1+(B/B^*_1)^{-3/2}]^{-1}-A_2B^{1/2}$ for Resistor A.  
}  
\label{fig:deltaR}
\end{figure}
%%%%%%%%%%%%%%%%%%

\section{Results and discussion}
\subsection{Magnetoresistance}
\begin{figure}
\centerline{
\psfig{file=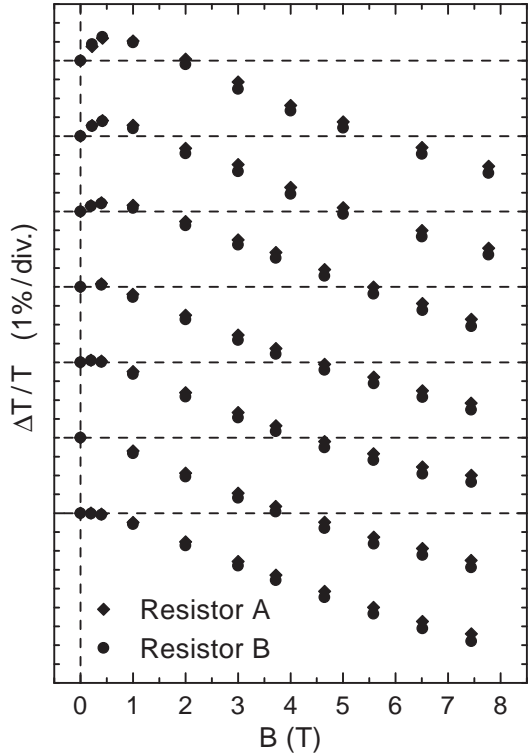,width=.8\columnwidth,
bbllx=69pt,
bblly=134pt,
bburx=509pt,
bbury=784pt,clip=,
angle=0}
}
\caption{
Relative change of apparent temperature as a function 
of magnetic field at constant temperatures.  
The temperatures are the same as in Figure~\ref{fig:deltaR}.  
The origin of each data set is offset for clarity.  
}  
\label{fig:deltaT}
\end{figure}

The relative change of resistance $\Delta R/R$ and 
of the corresponding apparent temperature $\Delta T/T$, 
i.e., temperature evaluated based on the $R-T$ calibration table 
for $B=0$, are shown in Figures~\ref{fig:deltaR} and \ref{fig:deltaT}, 
respectively, as functions of applied magnetic field for Resistors~A and B.
The two resistors are nominally identical (Note that their serial numbers 
are 1848 and 1849.) and the resistance at $B=0$ agrees within 1\% 
at all the temperatures of the magnetoresistance measurements.  
Hence, the orientation dependence of the magnetoresistance 
can be probed by comparing the results of the two resistors.  
One sees in Figures~\ref{fig:deltaR} and \ref{fig:deltaT} 
that the orientation dependence is much smaller than the magnitude 
of the magnetoresistance. 
This characteristic is preferable from a practical point of view 
because the users do not have to be worried about the orientation.  
In usual case where the canned chip is used, one cannot even know 
the orientation of the film.

%%%%%%%%%%%%%%%%%%%%%%%%%%%%%%%%%%%%%%%%%%%%%
%%%%  Fig.2: deltaT             %%%%%%%%%%%%%
%%%%%%%%%%%%%%%%%%%%%%%%%%%%%%%%%%%%%%%%%%%%%
%%%%%%%%%%%%%%%%%%%%%%%%%%%%%%%%%%%%%%%%%%%%%%

%%%%%%%%%%%%%%%%%%%%%%%%%
%%% Table I %%%
%%%%%%%%%%%%%%%%%%%%%%%%%
\begin{table}
\caption{Magnetic-field-induced relative resistance change 
$[R(B)-R(0)]/R(0)$ in units of \% at $B=0.5$~T.}  
\label{tab:nega}
%\begin{center}
\vspace*{2mm}
\begin{tabular}{cccccc} 
Resistor & 0.05 K & 0.06 K & 0.09 K & 0.12 K & 0.16 K \\
\hline
A & -1.7 & -1.1 & -0.6 & -0.2 & 0.0 \\
B & -1.8 & -1.0 & -0.5 & -0.1 & 0.0 \\
\end{tabular}
%\end{center}
\end{table} 
%%%%%%%%%%%%%%%%%%%%%%%%%

%%%%%%%%%%%%%%%%%%%%%%%%%
%%% Table II %%%
%%%%%%%%%%%%%%%%%%%%%%%%%
\begin{table}
\caption{Magnetic-field-induced relative resistance change 
$[R(B)-R(0)]/R(0)$ 
in units of \% at higher magnetic fields.  
The results for Resistors~A and B are compared 
with the values in Ref.~\ref{SIhttp}.}  
\label{tab:MR}
%\begin{center}
\vspace*{2mm}
\begin{tabular}{ccccccccccccc} 
&& \multicolumn{3}{c}{0.08 K} && \multicolumn{3}{c}{0.14 K}
&& \multicolumn{3}{c}{0.28 K} \\ 
    && A & B & Ref. \ref{SIhttp} && A & B & Ref. \ref{SIhttp} 
&& A & B & Ref. \ref{SIhttp} \\
\hline 
2 T && 1.0 & 1.3   & 1.4 && 1.5 & 1.8   & 1.3 && 1.3 & 1.4   & 1 \\
4 T && 3.6 & 4.1   & 4   && 3.7 & 4.0   & 3.4 && 3.1 & 3.3   & 3 \\
6 T && 6.1 & 6.6   & 7   && 5.6 & 6.0   & 5.3 && 4.5 & 4.7   & 5 \\
8 T && 8.2 & 8.7   & 8   && 7.3 & 7.7   & 7   && 5.6 & 5.9   & 6 \\
\end{tabular}
%\end{center}
\end{table} 
%%%%%%%%%%%%%%%%%%%%%%%%%%%%%%

Now we shall look at the sign and the magnitude of the magnetoresistance.  
Negative magnetoresistance appears at low-field regime at $T<0.2$~K, 
which is consistent with the information by SI~\cite{SIhttp} 
referring to the existence of an initial small negative magnetoresistance 
at temperatures below 0.25~K at low fields (less than about 1.5~T).  
The magnitude of the negative magnetoresistance, which is not 
described in Ref.~\ref{SIhttp}, grows rapidly 
as the temperature is lowered, and at $T=0.05$~K for example, 
$\Delta R(B)/R(0)$ at $B\approx0.5$~T is $-2\%$, 
which is no longer ``small".  
In Table~\ref{tab:nega} is given $\Delta R(B)/R(0)$ 
at $B=0.5$~T at several temperatures.  
When the RuO$_2$-based resistors are used as thermometers, the relatively 
strong temperature dependence of the magnetoresistance reduce the accuracy 
of the temperature measurements, and hence, an appropriate care should be 
taken with the negative magnetoresistance.  
The magnetic-field dependence at higher fields, 
on the other hand, is simple.  
In Table~\ref{tab:MR}, we summarize the magnetoresistance ratio 
of Resistors~A and B determined by the experiment at high fields.
The values given in the specifications~\cite{SIhttp} are also 
shown in the table.
We find the magnetoresistance at high fields are described 
well by the values in the specifications.

%%%%%%%%%%%%%%%%%%%%%%%%%%%%%%%%%%%%%%%%%%%%%
%%%%  Fig.3: param              %%%%%%%%%%%%%
%%%%%%%%%%%%%%%%%%%%%%%%%%%%%%%%%%%%%%%%%%%%%
\begin{figure}
\centerline{
\psfig{file=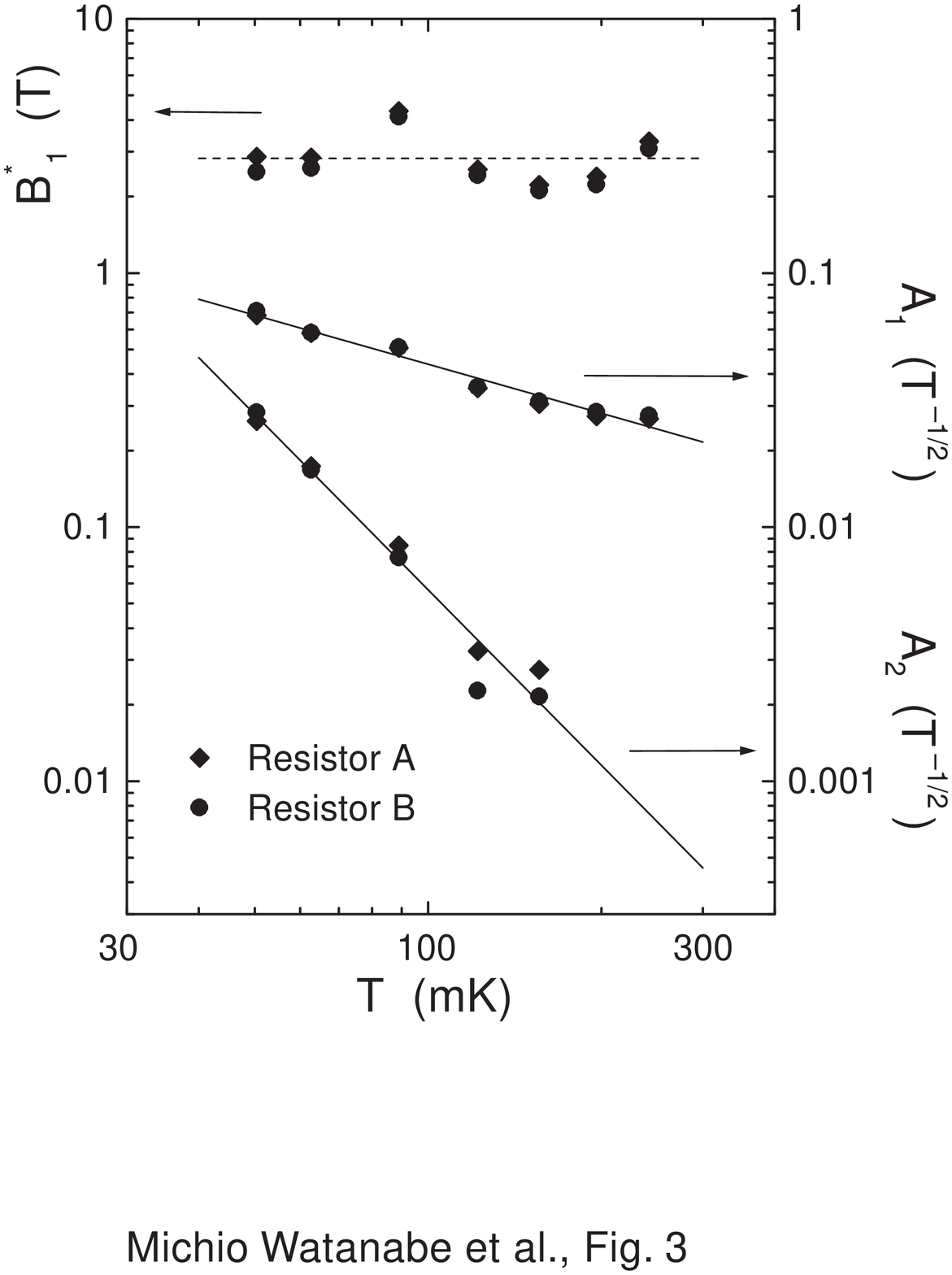,width=\columnwidth,
bbllx=37pt,
bblly=151pt,
bburx=571pt,
bbury=734pt,clip=,
angle=0}
}
\caption{
Parameters $A_1$, $A_2$, and $B^*_1$ obtained 
by fitting $r=r_1+r_2'$ 
[See Equations~(\ref{eq:r}), (\ref{eq:r1}), and (\ref{eq:r2'}).]  
vs. temperature.  
The horizontal dotted line represents the average of $B^*_2$, 
and the solid lines the best power-law fits for $A_1$ and $A_2$.  
}  
\label{fig:param}
\end{figure}
%%%%%%%%%%%%%%%%%%%%%%%%%%%%%%%%%%

\begin{figure}
\centerline{
\psfig{file=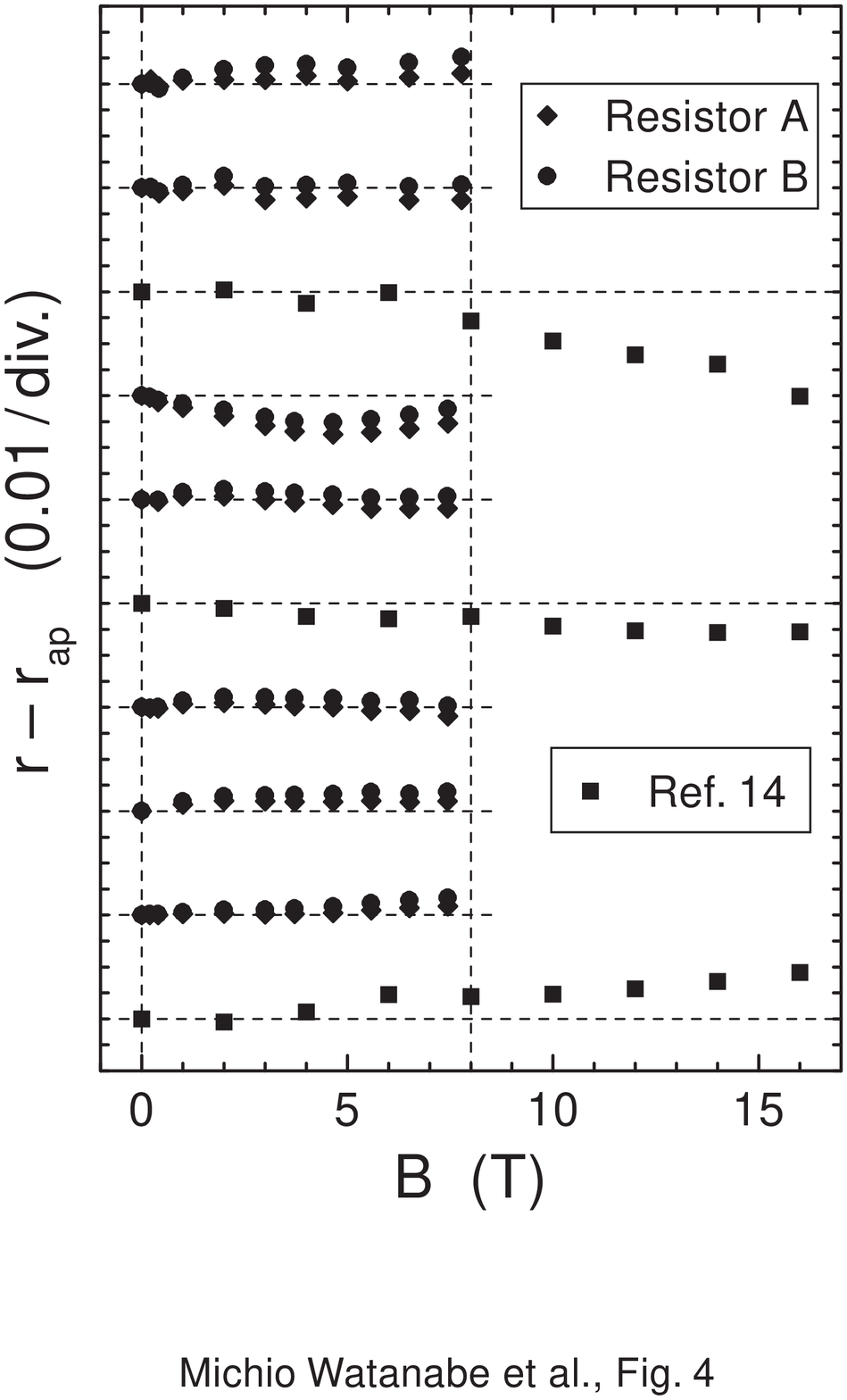,width=.8\columnwidth,
bbllx=56pt,
bblly=116pt,
bburx=511pt,
bbury=783pt,clip=,
angle=0}
}
\caption{
Deviation of the relative resistance change 
$r\equiv[R(B)-R(0)]/R(0)$ from $r_{ap}$ given 
by Equation~(\ref{eq:rap}). 
The temperatures from top to bottom in units of K are 0.05, 0.06, 
0.08, 0.09, 0.12, 0.14, 0.16, 0.20, 0.24, 0.28, respectively.  
The origin of each data set is offset for clarity.  
}  
\label{fig:devi}
\end{figure}

As an empirical expression for 
\begin{equation}
r\equiv [R(B)-R(0)]/R(0) 
\label{eq:r}
\end{equation}
of the SI resistors at $0.05\;{\rm K}\leq T < 0.3\;{\rm K}$, 
we propose 
\begin{eqnarray}
\nonumber 
r_{ap}(B,T)&\equiv&0.838\,T^{-0.641}\,B^{1/2}\left[1+\left(
\frac{B}{\,2.83\,}\right)^{\!\!-3/2}\;\right]^{-1}\\
&-&220\,T^{-2.29}\,B^{1/2}, 
\label{eq:rap}
\end{eqnarray}
where $B$ and $T$ are in units of T and mK, respectively.  
The underlying idea is as follows.  
Goodrich {\it et al.} measured the magnetoresistance 
up to 18~T or 32~T and reported $B^{1/2}$ 
dependence at $B>2.5-3$~T~\cite{Goo98}.  
In sufficiently low magnetic fields, on the other hand, 
magnetoresistance should show $B^2$ dependence 
irrespective of its origin 
so long as the symmetric relation $r(-B)=r(B)$ holds. 
Based on these points, we introduce 
\begin{equation}
\label{eq:r1}
r_1(B)\equiv A_1B^{1/2}\left[1+\left(\frac{B}{\,B^*_1\,}
\right)^{\!\!-3/2}\;\right]^{-1}, 
\end{equation}
where $A_1>0$ is a coefficient and $B^*_1$ is the magnetic field 
characterizing the crossover of the field dependence.  
Note that $r_1\approx (A_1B_1^{*\,-3/2})B^2$ for $0<B/B^*_1\ll1$ 
and $r_1\approx A_1B^{1/2}$ for $B/B^*_1\gg1$.  
We fit $r=r_1$ to the data in Figure~\ref{fig:deltaR} 
at $T=0.20$~K and 0.24~K, where only positive magnetoresistance 
is seen within the resolution of our measurements.   
The results for Resistor~A is shown in the same figure.  
In order to express the negative magnetoresistance 
which is clearly seen at lower temperatures, we add 
\begin{equation}
r_2(B)\equiv-A_2B^{1/2}\left[1+\left(\frac{B}{\,B^*_2\,}
\right)^{\!\!-3/2}\;\right]^{-1},
\end{equation}
where $0<A_2<A_1$ and $0<B^*_2<B^*_1$.  
In our resisters $B^*_2\ll0.1$~T is found, and hence 
\begin{equation}
\label{eq:r2'}
r_2\approx r_2'\equiv -A_2B^{1/2}
\end{equation}
is a good approximation at $B\geq0.1$~T.  
The curves in Figure~\ref{fig:deltaR} for Resistor~A 
at $T\leq0.16$~K are the fits of $r=r_1+r_2'$.  
The fits yield the values shown in Figure~\ref{fig:param}.  
The temperature dependences of $A_1$ and $A_2$ 
are described by the power law.  
This is the background of Equation~(\ref{eq:rap}).  
We have neglected the temperature variation of $B^*_1$ 
and used the average value 2.83~T. 
The deviation from $r_{ap}$ is shown in Figure~\ref{fig:devi}.  
The data from Ref.~\ref{SIhttp} are also plotted.  
All the data points at $B\leq8$~T 
are reproduced by Equation~(\ref{eq:rap}) within the error of 0.015.   

%%%%%%%%%%%%%%%%%%%%%%%%%%%%%%%%%%%%%%%%%%%%%
%%%%  Fig.4: devi               %%%%%%%%%%%%%
%%%%%%%%%%%%%%%%%%%%%%%%%%%%%%%%%%%%%%%%%%%%%
%%%%%%%%%%%%%%%%%%%%%%%%%%%%%%%%%%%%%%%%%%%%%

\subsection{Temperature dependence in zero magnetic field}
\begin{figure}
\centerline{
\psfig{file=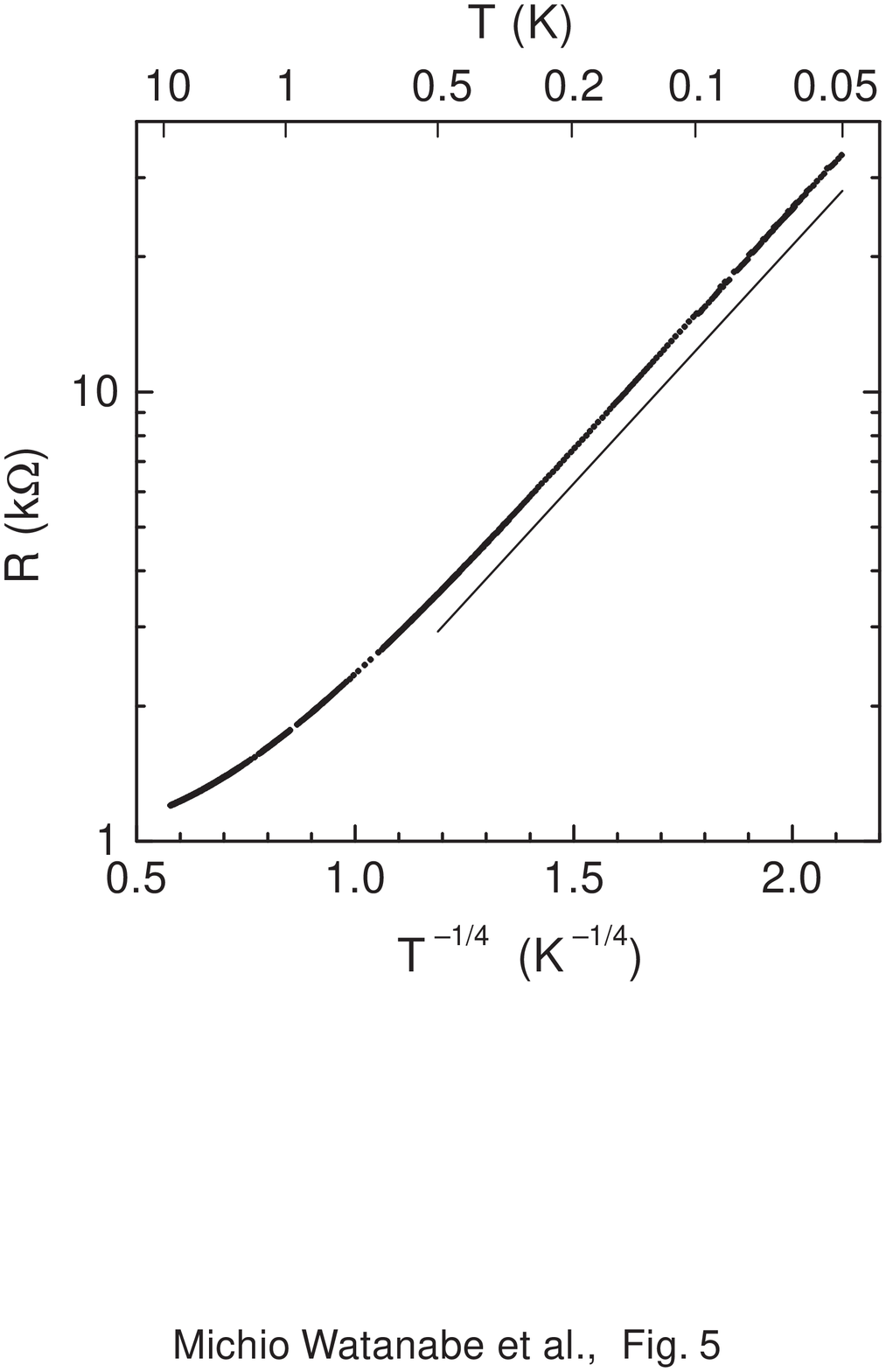,width=.85\columnwidth,
bbllx=33pt,
bblly=246pt,
bburx=508pt,
bbury=787pt,clip=,
angle=0}
}
\caption{
Resistance in zero magnetic field as a function of $T^{-1/4}$ 
for Resistor~B.  The line represents the best fit using the data 
between 0.05~K and 0.5~K.  
The fit is shifted downward for easier comparison.  
}  
\label{fig:Tdep}
\end{figure}

In Figure~\ref{fig:Tdep} the resistance in zero magnetic field 
is plotted as a function of $T^{-1/4}$ in a semi-log scale. 
We find the temperature variation of the resistance 
for Resistor~B at the lowest temperature range is 
described by 
\begin{equation} 
\label{eq:VRH}
R(T)\propto\exp[(T_0/T)^p].  
\end{equation}
with $p=1/4$.  
The temperature dependence of the resistance 
of RuO$_2$-based resistors at low temperatures has been analyzed 
in terms of variable-range hopping (VRH) 
conduction~\cite{Doi84,Li86,Wil90,Bri91,Nep96,Bat95}.  
In some works~\cite{Wil90,Bri91,Bat95}, 
the exponent $p$ was treated as a fitting parameter 
and various values ($0.14\leq p<0.71$) were reported.  
According to the theory of VRH~\cite{Shk84}, $p$ is determined 
by the dimensionality $d$ and the shape of the density of states 
around the Fermi level.  
For $d=3$ and the single-particle density of states $g(E)$ expressed by
\begin{equation}
g(E)\propto|E-E_F|^n, 
\end{equation}
where $E_F$ is the Fermi energy, the exponent $p$ is given by 
\begin{equation}
p = \frac{n+1}{\,n+4\,}\,.
\end{equation}
For {\em thick}-film chip resistors $d=3$ is reasonable, and 
the very small orientation dependence of the magnetoresistance 
that we saw in the preceding subsection is consistent with this idea.  
The present result, $p=1/4$, supports the constant density of states, $n=0$, 
around the Fermi level rather than the Coulomb pseudo gap, $n=2$. 
Good fits with $p=1/4$ have also been reported for 
the resistors manufactured both by ALPS~\cite{Doi84} 
and by Dale (RCWP575~\cite{Li86}, RC550~\cite{Nep96}).  

%%%%%%%%%%%%%%%%%%%%%%%%%%%%%%%%%%%%%%%%%%%%%
%%%%  Fig.5: Tdep               %%%%%%%%%%%%%
%%%%%%%%%%%%%%%%%%%%%%%%%%%%%%%%%%%%%%%%%%%%%
%%%%%%%%%%%%%%%%%%%%%%%%%%%%%%%%%%%%%%%%%%%%%%

\subsection{Applicability below 0.05~K}
At very low temperatures, resistors are 
easily overheated, i.e., the measured temperature can be considerably 
higher than the real temperature.  
An obvious source of the overheating is the Joule heat $P=RI^2$ 
by the bias current $I$ for the resistance measurements. 
Besides $P$, however, there can be other sources of overheating, 
such as the energy dissipation due to the high-frequency noise, 
due to the induced current by a ground loop, due to the thermoelectric 
power, and so on. 
Heat conduction through the electrical leads can be 
other source of heat input in case of improper thermal anchoring. 
Because of these, it is not an easy task to ensure the absence 
of overheating in experiments.  
Fortunately, in the present case, we can determine 
whether overheating is present or not, 
because the temperature dependence of the resistance given by
Equation~(\ref{eq:VRH}) with $p=1/4$ is reasonably expected to 
hold even at temperatures lower than 0.05~K. 

\begin{figure}
\centerline{
\psfig{file=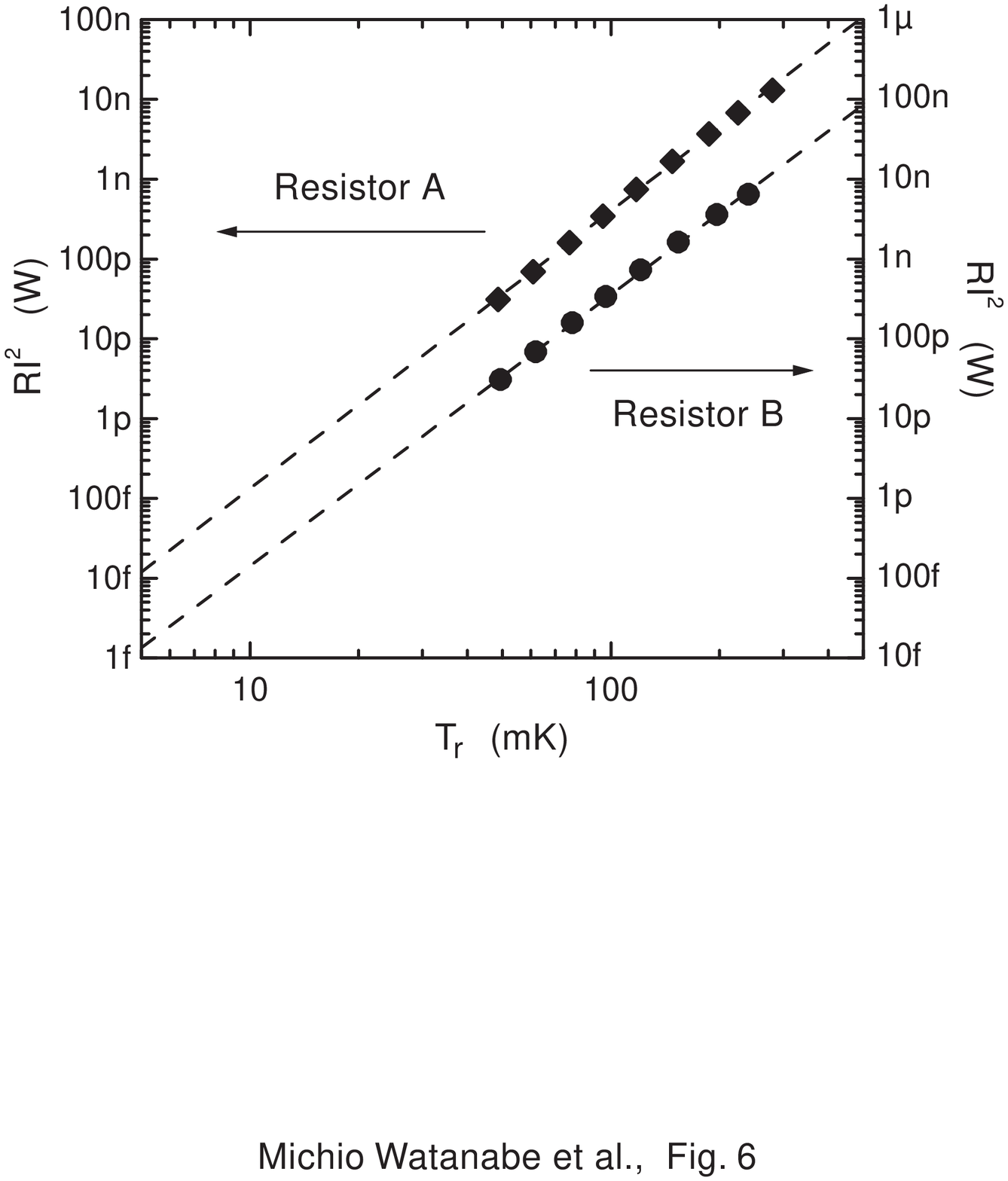,width=\columnwidth,
bbllx=26pt,
bblly=277pt,
bburx=575pt,
bbury=700pt,clip=,
angle=0}
}
\caption{
Measuring power vs. the temperature of the resistor 
determined from its resistance.  
During the measurements, the temperature 
of the mixing chamber was lower than 6~mK.  
}  
\label{fig:heat}
\end{figure}

In order to estimate how small the heat flow into the resistors 
should be, we measured the resistance at various bias currents 
in zero magnetic field.  
In Figure~\ref{fig:heat} we show $P$ vs. the temperature 
of the resistor $T_r$ determined from its resistance.  
During the measurements, the temperature of the mixing chamber 
$T_m$ was always lower than 6~mK.  
All the data points align on straight lines, i.e., 
\begin{equation}
\label{eq:bigP}
\dot{Q}\approx AT_r^{\,q} 
\end{equation}
holds, where $\dot{Q}$ is the heat input to the resistor, 
which must be dominated by $P$ when $T_r\gg T_m$ 
as in Figure~\ref{fig:heat}, and $A$ is a coefficient.  
The dashed lines in Figure~\ref{fig:heat} represent fits of 
Equation~(\ref{eq:bigP}).  
We obtain $q=3.5\pm0.1$ and $\log_{10} A=-5.9\pm0.02$ for Resistor~A, 
and $q=3.4\pm0.1$ and $\log_{10} A=-6.0\pm0.03$ for Resistor~B, 
where $A$ is in units of W/K$^q$.   

%%%%%%%%%%%%%%%%%%%%%%%%%%%%%%%%%%%%%%%%%%%%%
%%%%  Fig.6: heat               %%%%%%%%%%%%%
%%%%%%%%%%%%%%%%%%%%%%%%%%%%%%%%%%%%%%%%%%%%%
%%%%%%%%%%%%%%%%%%%%%%%%%%%%%%%%%%%%%%%%%%%%%%

Although the lowest temperature in Figure~\ref{fig:heat}, 
$T_r=49$~mK, is much higher than $T_m$, we expect that the dependence 
of Equation~(\ref{eq:bigP}) holds even for much smaller heat input. 
In this case, we should modify the equation as   
\begin{equation} 
\label{eq:Qdot}
\dot{Q}=A(T_r^{\,q}-T_m^{\,q}) 
\end{equation}
because $T_r$ should be equal to $T_m$ when $\dot{Q}=0$.
We should note that this relationship is expected when there is 
a distinct bottleneck in the path of heat flow from electron system 
in the sensor to the refrigerator. 
Possible sources of such bottleneck are poor electron-phonon coupling 
in the RuO$_2$, the Kapitza resistance, and the thermal resistance of 
electrically insulating material.  
Unfortunately, we cannot identify which is the case at present. 
If Equation~(\ref{eq:Qdot}) holds for arbitrary $T_m$ and $T_r$, 
it means that the thermal impedance 
\begin{equation}
Z=\lim_{\dot{Q}\to0}\frac{\,\Delta T\,}{\dot{Q}}
\end{equation}
between the electron system and the refrigerator is proportional 
to $T^{-q+1}$, where $\Delta T\equiv T_r-T_m$.  
Such power-law dependence of $Z$ on $T$ has been reported 
for RuO$_2$-based resistors~\cite{Wil90}. 
 
\begin{figure}
\centerline{
\psfig{file=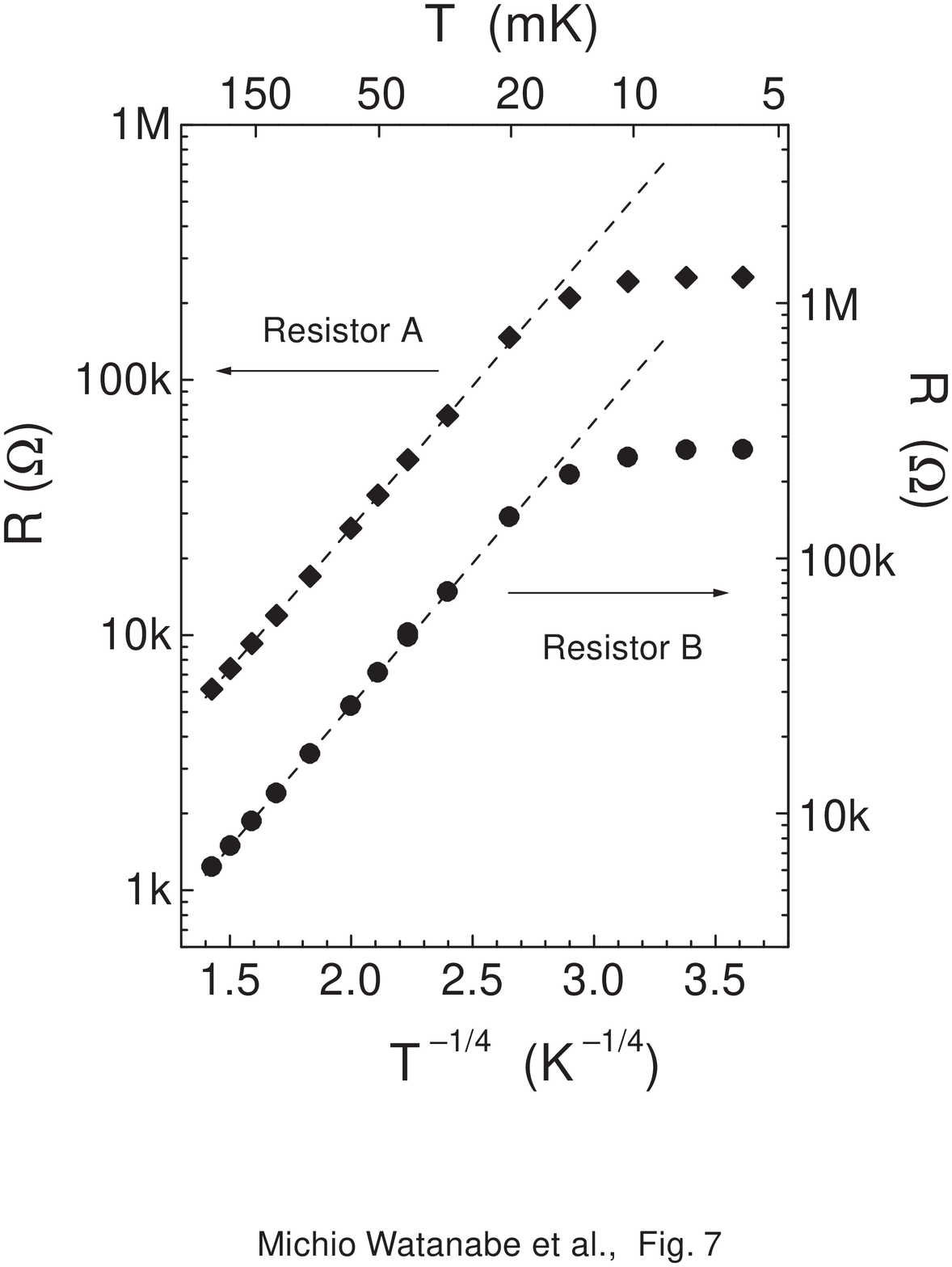,width=\columnwidth,
bbllx=20pt,
bblly=150pt,
bburx=576pt,
bbury=799pt,clip=,
angle=0}
}

\caption{
Resistance as a function of $T^{-1/4}$ for both the resistors.  
}  
\label{fig:Mott}
\end{figure}

Using an approximation of Equation~(\ref{eq:Qdot}) 
in the limit of $\Delta T \ll T_m$, 
\begin{equation} 
\label{eq:equi}
\dot{Q}\approx Aq(\Delta T/T_r)T_r^{\,q}, 
\end{equation}
we can evaluate the maximum value allowed for $\dot{Q}$, 
which is simultaneously the upper bound for $P$, 
for a given accuracy in temperature determination, $\Delta T/T_r$.  
For example, in order to achieve $\Delta T/T_r<0.01$,  
$P<0.2$~pW is required at 30~mK and $P<5$~fW at 10~mK.  
Keeping $P$ sufficiently low, we measured the resistance down to 
6~mK as shown in Figure~\ref{fig:Mott}.  For $T>15$~mK the resistance is 
described well by Equation~(\ref{eq:VRH}) with $p=1/4$, 
and hence the resistors are applicable.  
The temperature dependence of resistance levels off below 15~mK 
in spite of small excitation current. 
This leveling-off is explained if we assume the excess heat input 
to the resistor of about 0.1~pW.  

%%%%%%%%%%%%%%%%%%%%%%%%%%%%%%%%%%%%%%%%%%%%%
%%%%  Fig.7: Mott               %%%%%%%%%%%%%
%%%%%%%%%%%%%%%%%%%%%%%%%%%%%%%%%%%%%%%%%%%%%
%
%%%%%%%%%%%%%%%%%%%%%%%%%%%%%%%%%%%%%%%%%%%%%

\section{Conclusion}
We have determined the magnetoresistance of RuO$_2$-based resistors 
produced by Scientific Instruments Inc. 
between 0.05~K and 0.3~K in magnetic fields up to 8~T.  
Negative magnetoresistance is observed around 0.5~T, 
and its magnitude grows rapidly as the temperature is lowered.   
Positive magnetoresistance at high magnetic fields 
has smaller temperature dependence, and its magnitude is consistent 
with the information provided by the company.  
We have also investigated the characteristics of the resistors 
below 50~mK in zero magnetic field.  
The resistors are applicable down to at least 15~mK.  

\section*{Acknowledgements}
The measurements were carried out 
at the Cryogenic Center, University of Tsukuba, Japan. 
M. W. would like to thank Japan Society for the Promotion 
of Science (JSPS) for financial support.

\end{multicols}
\end{document}